\documentclass[manuscript]{aastex}

\usepackage{graphicx}% Include figure files
\newcommand{\sss}{\scriptscriptstyle}
\newcommand {\be}{\begin{equation}} % start equation
\newcommand{\ee}{\end{equation}}    % end equation
\def\vtj{v_{{\sss T}j}}
\def\vti{v_{{\sss T}i}}

\def\vte{v_{{\sss T}e}}

\shorttitle{Kinetic instability of drift-Alfv\'{e}n waves} \shortauthors{Vranjes and Poedts}

\begin{document}

\title{Kinetic instability of drift-Alfv\'{e}n waves in solar corona and stochastic heating}


\author{J. Vranjes and S. Poedts%\altaffilmark{1}
} \affil{Center for Plasma Astrophysics, \\ and Leuven Mathematical Modeling and Computational Science Center (LMCC),\\
K. U. Leuven, Celestijnenlaan 200B, 3001 Leuven,  Belgium. \email{Jovo.Vranjes$@$wis.kuleuven.be;
Stefaan.Poedts$@$wis.kuleuven.be}}

\begin{abstract}
The solar atmosphere is structured and inhomogeneous both horizontally and vertically. The omnipresence of coronal
magnetic loops implies
 gradients of the equilibrium plasma quantities like the density, magnetic field and temperature. These gradients are responsible for the
 excitation of drift waves that grow both within the two-component fluid description (in the presence of collisions and without it) and within the
 two-component kinetic descriptions (due to purely kinetic effects). In the present work the effects of the density gradient in the
 direction perpendicular to the magnetic field vector are investigated within the kinetic theory, in both electrostatic and
 electromagnetic regimes. The electromagnetic regime implies the coupling of the gradient-driven drift wave with the Alfv\'{e}n wave.
The growth rates for the two cases are calculated and compared. It is found that, in general, the electrostatic regime
is characterized by stronger growth rates, as compared with the electromagnetic perturbations. Also discussed is  the
stochastic heating associated with the drift wave.  The released amount of energy density due to this heating  should
be more dependent on the magnitude of the background magnetic field than on the coupling of the drift and Alfv\'{e}n
waves. The stochastic
 heating is  expected to be  much higher in regions with a stronger  magnetic field. On the whole,  the energy release
 rate  caused by the stochastic heating   can be several orders of magnitude above the value presently accepted  as
  necessary for a sustainable coronal heating. The vertical stratification and the very long wavelengths along the
   magnetic loops imply that a drift-Alfv\'{e}n wave, propagating as a twisted structure along the loop, in fact
   occupies regions with different plasma-$\beta$ and, therefore, may have different (electromagnetic-electrostatic)
    properties, resulting in different heating rates within just one or two wavelengths.
\end{abstract}


\section{Introduction}

Observations and theoretical studies in the past 70 years have dramatically increased our knowledge and understanding
of the physical processes in the solar atmosphere. However, the basic starting puzzle of the problem of coronal heating
still remains elusive in spite of the obvious progress made in the domain. In fact, new data collected in the course of
decades have additionally increased the complexity of the problem as more and more fine details related to the heating
have emerged. These include the preferential heating of plasma particles in the direction perpendicular to the magnetic
field vector, resulting in a temperature anisotropy \citep{li,cus}, and the preferential heating of heavier particles.
As a matter of fact, heavier ions appear to be hotter than lighter ions, the latter on the other hand appear to be
hotter than electrons \citep{cr3}. Moreover, extremely strong electric fields (above $100\;$ kV/m) have been detected
\citep{zs}. Those electric fields accelerate particles and, in general, the distribution functions of the plasma
species in the outer solar atmosphere can considerably be different from a Maxwellian distribution \citep{vas,cr}.

 In
our recent papers \citep{v1, v2, v3, v4} a novel approach and a new paradigm for the coronal heating has been put
forward. The model is based on the drift wave theory, a well known subject in the general  plasma theory, in laboratory
plasma physics, and even in  terrestrial ionospheric research \cite{kel}. Yet, the drift wave theory is completely
overlooked in the context of solar plasmas. It implies the abundance of free energy for the instability of the drift
wave already in the corona. That energy is  stored in the gradients of the density, temperature and magnetic field. It
makes the waves growing and  it results in heating due to the polarization drift effects. The heating is stochastic by
nature; for short enough wave-lengths in the direction perpendicular to the magnetic field vector, gyrating plasma
particles feel a space-time  variation of the wave-electric field, and their motion becomes stochastic and equivalent
to the increase of the temperature. The nature of the heating  is such that it acts essentially in the perpendicular
direction, and,  in addition to this, more massive particles are in fact more effectively heated by that mechanism. The
analysis performed in \citet{v1, v2, v3, v4} was focused on the electrostatic domain of the drift wave instability.
This implies a small plasma-$\beta= 2\mu_0 n_0 \kappa T/B_0^2$, e.g.\ of the order or below $m_e/m_i$. However, even in
that domain the plasma may support electromagnetic (EM) perturbations too \citep{kral}, although as a rule those will
not be well coupled to the electrostatic ones. For a plasma-$\beta$ value above $m_e/m_i$, the coupling will
effectively take place and, as a first manifestation of this, only the bending of the magnetic field vector may be
taken into account. Such a coupled drift-Alfv\'{e}n mode has in fact  been studied in our earlier work \citet{vaa} by
using two-component fluid theory, with a complete and self-consistent  contribution  of hot ions effects, appropriate
for  the hot solar corona. Within such a  two-fluid theory, the  drift-Alfv\'{e}n mode is destabilized in the presence
of collisions and, for a large enough parallel wave-number, the mode has all the features of a growing  Alfv\'{e}n
wave. This is because of an exchange of identities of the drift and Alfv\'{e}n modes \citep{w,vaa} occurring in a
certain parameter domain.

On the other hand, the drift-Alfv\'{e}n wave instability within the collision-less kinetic theory has a completely
different nature, and this will be the subject of the present work. In particular, we shall investigate the difference
in the growth-rates of the electrostatic (drift) and the electromagnetic (drift-Alfv\'{e}n) modes. Such a difference is
expected because of the following two opposite effects: i)~a part of the energy that drives the instability is spent on
the bending of the magnetic field vector and this should in principle reduce the growth rate, yet in the same time,
ii)~this bending should partly reduce the electron mobility in the parallel direction. The effects of such a  reduction
should be  similar to electron collisions studied by  \citet{vaa} and, as a result, the growth-rate may become
increased. Hence, the total outcome of the EM effects will then depend on the mutual  ratio of these two opposite
effects. A local analysis will be used. This is very appropriate for a geometry without magnetic shear, and for the
case in which the perpendicular component of the wave-length is much shorter than the characteristic lengths of the
equilibrium gradients   \citep{kral}.

\section{Basic equations}
In the case of a plasma beta in the range $m_e/m_i\leq\beta\ll 1$,  it is appropriate to take  into account only the
bending of the magnetic field. The perturbed density for the species $j=e, i$ is, in that case,  described by
\citep{w}:
\[
n_{j1}= - \frac{q_j n_{j0}}{\kappa T_j} \left\{\phi_1+ (\omega- \omega_1) \phi_1 \sum_m\frac{\Lambda_m(b_j)}{\omega_2-
m\Omega_j} \times \right.
\]
\[
\times \left. \left[W\left(\frac{\omega_2- m\Omega_j}{|k_z| \vtj}\right) - 1\right]\right.
\]
\be \left. + (\psi_1 -\phi_1) \left(1- \frac{\omega_1}{\omega}\right) \, \sum_m \Lambda_m(b_j) W\left(\frac{\omega_2-
m\Omega_j}{|k_z| \vtj}\right) \right\}. \label{e1} \ee
Here, $\omega_1=\omega_{*j} - k_y g/\Omega_j$, $\omega_2= \omega + k_y g/\Omega_j$,
  $\omega_{*j}= v_{*j} k_y$, where $\vec
v_{*j}=-\vec e_z\times \nabla p_{j0}/(q_j n_{j0} B_0)$ is the diamagnetic velocity. The equilibrium magnetic field and
density gradient are $\vec B_0=B_0 \vec e_z$ and $\nabla n_{j0}= -\vec e_x dn_{j0}/dx$, and we use a  local
approximation and Fourier analysis with  small  perturbations of the form $\sim \hat{f}(x)\exp(- i \omega t + i k_y y +
i k_z z)$, where $\hat{f}(x)$ is the $x$-dependent amplitude and $|d/dx|\ll |k_y|$.  In the terms  $\omega_{1,2}$ we
have the gravity effects that are  included  through the Maxwellian distribution function  with a $ m_jgx$ term in the
exponent. Below, it will
 be kept for ions only.  The other notation is as follows
  $b_j=k_y^2\rho_j^2$, $\Lambda_m(b_j)= I_m(b_j) \exp(-b_j)$,  $\rho_j=\vtj/\Omega_j$, $\vtj^2=\kappa T_j/m_j$, $\Omega_j=q_j B_0/m_j$,  $I_m$ is the modified Bessel function of the
first kind and order $m$,  $W(\chi)=(2 \pi)^{1/2} \int_{-\infty}^{+\infty} \eta \exp(- \eta^2/2) d\eta/(\eta- \chi)$.
The terms $\phi$ and $\psi$ describe the potential of the electric field \citep{w}, $\vec E=-\nabla_\bot \phi - \vec
e_z \partial \psi/\partial z$, and for that reason an additional equation for the parallel  current is needed in order
to have a closed set
\[
j_{jz1}=-\frac{q^2 n_{j0}}{\kappa T_j k_z} \left[ (\omega-\omega_1) \phi_1 \sum_m \Lambda_m(b_j) W\left(\frac{\omega_2-
m\Omega_j}{|k_z| \vtj}\right)\right.
\]
\[
\left. + (\psi_1-\phi_1) \left(1-\frac{\omega_1}{\omega}\right) \times \right.
\]
\be \left. \times \sum_m \Lambda_m(b_j) (\omega_2- m \Omega_j)  W\left(\frac{\omega_2- m\Omega_j}{|k_z|
\vtj}\right)\right]. \label{j} \ee
For electrons it is good enough to use  a negligible mass limit so that  $\Lambda_0(b_e)\simeq 1$, while the  deviation
from unity of the corresponding term for ions is a finite Larmor radius effect. In the limit of  frequencies
$|\omega_2|$ much below $\Omega_j$, one keeps only the term $m=0$ in the summation for both electrons and ions
    \citep{s,w,b}. For similar  reasons, in the case $|\chi|< 1$, we shall use the approximate expression (for electrons)  $W(\chi)\simeq i (\pi/2)^{1/2} \chi
\exp(-\chi^2/2) +1 -\chi^2 + \chi^4/4 \ldots$, and   for ions $|\chi|> 1$,  $W(\chi)\simeq  i (\pi/2)^{1/2} \chi
\exp(-\chi^2/2)-1/\chi^2 -3/\chi^4 +\ldots$. For these two species Eq.~(\ref{e1}) then becomes \citep{w}
\[
\frac{n_{e1}}{n_0}= \frac{e }{\kappa T_e}\left\{\phi_1 + i \phi_1 \left(\frac{\pi}{2}\right)^{1/2}
\frac{\omega-\omega_{*e}}{|k_z| \vte}\exp\left(-\frac{\omega^2}{2 k_z^2 \vte^2}\right)\right.
\]
\[
\left. + (\psi_1 -\phi_1) \left(1- \frac{\omega_{*e}}{\omega}\right) \left[1 + i \left(\frac{\pi}{2}\right)^{1/2}\times
\right. \right.
\]
\be \left. \left. \times \frac{\omega}{|k_z| \vte} \exp\left(-\frac{\omega^2}{2 k_z^2 \vte^2}\right) \right]\right\},
\label{e2} \ee
\[
\frac{n_{i1}}{n_0}= -\frac{e }{\kappa T_i}\left\{\phi_1 -  \phi_1 \frac{\omega-\omega_1}{\omega_2}\Lambda_0(b_i)
\left[1+ \frac{k_z^2 \vti^2}{\omega_2^2} +\frac{3k_z^4 \vti^4}{\omega_2^4}\right.\right.
\]
\[
\left. \left. - i \left(\frac{\pi}{2}\right)^{1/2} \frac{\omega_2}{|k_z| \vti} \exp\left(-\frac{\omega_2^2}{2 k_z^2
\vti^2}\right)\right] +(\psi_1 -\phi_1) \times \right.
\]
\[
\left. \times\left(1- \frac{\omega_1}{\omega}\right) \Lambda_0(b_i)  \left[ i \left(\frac{\pi}{2}\right)^{1/2}
\frac{\omega_2}{|k_z| \vti}\exp\left(-\frac{\omega_2^2}{2 k_z^2 \vti^2}\right) \right. \right.
\]
\be
\left.\left.
 - \frac{k_z^2 \vti^2}{\omega_2^2}-
\frac{3k_z^4 \vti^4}{\omega_2^4} \right]\right\}. \label{e3} \ee

\section{Electrostatic  drift wave  instability }
In the electrostatic limit, one may  set $\phi_1=\psi_1$ in the above expressions. Using   the  quasi-neutrality
condition  $n_i=n_e$ and Eqs.~(\ref{e2},\ref{e3}),  one  directly obtains
 the dispersion equation in the form $Re \Delta (\omega, \vec k) + i Im \Delta(\omega, \vec k) =0$. The frequency is assumed to be complex,  in the form $\omega=\omega_r + i \gamma$. Setting
 $Re \Delta (\omega_r, \vec k) =0$, one then obtains the following equation for the real part of the frequency:
\[
\omega_{2r}^5 \left(1+ \frac{T_e}{T_i}\right) -\Lambda_0(b_i) \frac{T_e}{T_i} \omega_r \omega_{2r}^4 + \Lambda_0(b_i)
\frac{T_e}{T_i}\omega_1\omega_{2r}^4
\]
\[
- \Lambda_0(b_i) k_z^2 c_s^2 \omega_r\omega_{2r}^2 + \Lambda_0(b_i)  \omega_1 k_z^2 c_s^2 \omega_{2r}^2 - 3
\Lambda_0(b_i) k_z^4 \vti^2 c_s^2 \omega_r
\]
\be + 3 \Lambda_0(b_i) k_z^4 \vti^2 c_s^2\omega_1=0. \label{s}
 \ee
In the limit of a negligible ion response along the magnetic field vector, $|\omega_r/k_z|\ll c_s$ with $c_s^2=\kappa
T_e/m_i$,  and for small gravity effects, this gives  the electrostatic drift wave frequency used in
 \citet{v1,v2,v3,v4}:
% \citet{v1,v2,v3}:
%
\be
\omega_r=-\frac{\omega_{*i} \Lambda_0(b_i)}{1- \Lambda_0(b_i) + T_i/T_e }= \frac{\omega_{*e} \Lambda_0(b_i)}{1+
[1-\Lambda_0(b_i)]T_e/T_i }\label{k1}
 \ee
Here,  $\omega_{*i}=-\omega_{*e} T_i/T_e$. Observe that after setting $\Lambda_0(b_i)\simeq 1- b_i$, the term in the
denominator becomes $1+ k_y^2 \rho_s^2$, where $\rho_s=c_s/\Omega_i$. Using the two-fluid description for comparison
and as a guideline \citep{b,vaa}, it can be shown that this  same expression (describing the finite ion inertia)
directly follows from the ion polarization drift, the latter playing an essential role in the process of stochastic
heating \citep{b1,b2} that will be discussed below.

On the other hand, for a purely parallel propagation from Eq.~(\ref{s}) one obtains the ion-acoustic (IA) mode in
plasmas with hot ions
 \be
 \omega_r^4- k_z^2 c_s^2 \omega_r^2 - 3 k_z^4 \vti^2 c_s^2=0. \label{ia}
 \ee
 The  solutions $\omega_r^2= (k_z^2 c_s^2/2) [1+ (1+ 12 T_i/T_e)^{1/2}]$  for $T_i=T_e$ yield the
  frequency $\omega_r\simeq \pm 1.5\, k_z c_s$. Hence, the ion temperature plays no important role in
  the real part of the IA wave frequency, while  the opposite is certainly true  with its imaginary part.

 The growth rate $\gamma$   is obtained approximately from  $\gamma \simeq -Im\Delta(\omega_r, \vec k)/[\partial Re\Delta/\partial \omega]_{\omega=\omega_r}$. This yields
 \[
 \gamma=-\left(\frac{\pi}{2}\right)^{1/2}\, \frac{\omega_{2r}^2}{\omega_{*e}\Lambda_0(b_i)} \left[\frac{\omega_r-\omega_{*e}}{|k_z|\vte} \, \exp\left(-\frac{\omega_r^2}{2 k_z^2 \vte^2}\right)
 \right.
 \]
 \[
 \left.
 +  \Lambda_0(b_i) \frac{T_e}{T_i} \frac{\omega_r-\omega_1}{|k_z|\vti} \, \exp\left(-\frac{\omega_{2r}^2}{2 k_z^2 \vti^2}\right)\right]\times
 \]
 \be
 \times \left[1+ \frac{k_z^2\vti^2}{\omega_{2r}^2}\left(3- \frac{2 \omega_{2r}}{\omega_{*i}}\right) + \frac{3 k_z^4\vti^4}{\omega_{2r}^4}\left(5- \frac{4 \omega_{2r}}{\omega_{*i}}\right)\right]^{-1}.
 \label{gr}
 \ee
The frequency on the right-hand side in Eq.~(\ref{gr}) is to be obtained from Eq.~(\ref{s}). Note that in the absence
of gravity  $\omega_{2r}\rightarrow \omega_r$, and in addition, for a negligible ion response along the magnetic field
vector, Eq.~(\ref{gr}) yields the growth rates from \citet{v1,v2,v3,v4}. For the assumed geometry $\omega_{*i}$ is
negative and the ion contribution will always tend to reduce  the growth rate, while electrons will make the mode
growing. Clearly this purely kinetic instability can only take place provided that  the frequency $\omega_r$ is  below
$\omega_{*e}$.

\section{Electromagnetic  perturbations}
To proceed with the electromagnetic perturbations, a  procedure similar to the derivation of  Eqs. (\ref{e2}) and
(\ref{e3}) yields the electron parallel current \be
j_{ez1}=-\frac{e^2 n_0}{k_z \kappa T_e} (\omega- \omega_{*e}) \left(1+ i \Upsilon_e\right) \psi_1, \label{e4} \ee
\[
\Upsilon_e= \left(\frac{\pi}{2}\right)^{1/2} \frac{\omega}{|k_z| \vte} \exp\left(-\frac{\omega^2}{2 k_z^2
\vte^2}\right).
\]
The corresponding expressions for the ions are:
\[
j_{iz1}=\frac{e^2 n_0}{k_z \kappa T_i} (\omega- \omega_1) \Lambda_0(b_i)  \left[\left(1+ \frac{\omega_g}{\omega}
\right)\psi_1 - \frac{\omega_g}{\omega} \phi_1\right] \times
\]
\be
\times \left[\frac{k_z^2 \vti^2}{\omega_2^2} \left(1+ \frac{k_z^2 \vti^2}{\omega_2^2}\right) -   i \Upsilon_i\right],
\label{e4a} \ee
\[
\Upsilon_i= \left(\frac{\pi}{2}\right)^{1/2} \frac{\omega_2}{|k_z| \vti} \exp\left(-\frac{\omega_2^2}{2 k_z^2
\vti^2}\right), \quad \omega_g=\frac{k_y g}{\Omega_i}.
\]
The first necessary equation is obtained as above,  by using the quasi-neutrality condition $n_{i1}=n_{e1}$ and
Eqs.~(\ref{e2} and \ref{e3}). The second equation follows from  the Amp\`{e}re law that, with the help of
Eqs.~(\ref{e4}) and (\ref{e4a}),  yields \be
\psi_1=\phi_1 \, \frac{s_1+ f_1}{s_2+ f_2}, \label{e5} \ee
\[
s_1=k_y^2 \rho_s^2 k_z^2 c_a^2, \quad s_2= s_1+ \omega(\omega_{*e} - \omega) \left(1+ i \Upsilon_e\right),
\]
\[
f_1=\frac{T_e}{T_i} \Lambda_0(b_i) \omega_g (\omega-\omega_1) \left(\alpha- i\Upsilon_i\right), \quad f_2=f_1\,
\frac{\omega_2}{\omega_g},
\]
\[
 \alpha=\frac{k_z^2\vti^2}{\omega_2^2} \left(1+\frac{k_z^2\vti^2}{\omega_2^2}\right).
\]
Here, $c_a^2=B_0/(\mu_0 n_0 m_i)$. The terms $f_{1,2}$ originate from the ion parallel current and in many situations
can be neglected.

The potential (\ref{e5}) is used to eliminate $\psi_1$  in the quasi-neutrality condition and in the end one obtains
the following dispersion  equation:
\[
-\frac{T_e}{T_i}\left[1- \frac{\omega- \omega_1}{\omega_2} \Lambda_0(b_i) \left(1+ \alpha - i \Upsilon_i\right)\right.
\]
\[
\left.
 + p\,\left(\frac{s_1+ f_1}{s_2+ f_2} - 1\right)\frac{\omega- \omega_1}{\omega} \Lambda_0(b_i) \left(i\Upsilon_i - \alpha\right) \right]
 \]
 \be
=1+ i\Upsilon_* + p\,\left(\frac{s_1+ f_1}{s_2+ f_2} - 1\right)\left(1-\frac{\omega_{*e}}{\omega}\right) \left(1+ i
\Upsilon_e\right). \label{e6} \ee Here,
\[
 \Upsilon_*=\left(\frac{\pi}{2}\right)^{1/2}
\frac{\omega -\omega_{*e}}{|k_z| \vte} \exp\left(-\frac{\omega^2}{2 k_z^2 \vte^2}\right).
\]
The parameter $p$ is set here by hand, and for convenience only; taking $p=0$ is equivalent to the electrostatic limit
discussed in the previous section, while $p=1$ implies the electromagnetic perturbations that are of interest here.

The dispersion equation (\ref{e6}) describes the coupled Alfv\'{e}n and drift waves as well as the ion parallel
(acoustic) response, together with the gravity and finite Larmor radius  effects  for ions. Note that, for the
parameters used further in the text, the gravity drift frequency $k_y g/\Omega_i$ is usually negligible.

The wave spectra are obtained following the same procedure  as before yielding from the real part of Eq.~(\ref{e6}):
\[
0=1+ \frac{T_e}{T_i}\left[ 1- \frac{\omega_r- \omega_1}{\omega_{2r}} \Lambda_0(b_i) (1+ \alpha_r)\right]
\]
\be +  \left(\frac{s_{1r}+ f_{1r}}{s_{2r}+ f_{2r}}-1\right) \left[1-\frac{\omega_{*e}}{\omega_r}  - \Lambda_0(b_i)
\alpha_r \frac{T_e}{T_i}  \left(1-\frac{\omega_1}{\omega_r}\right)\right]. \label{e7} \ee Here, the index $r$ denotes
the real part of the corresponding expressions.

It is interesting to compare these derivations with the results from the two-component fluid theory. Omitting the ion
parallel response and gravity, Eq.~(\ref{e7}) yields
%\[
%\omega_r^3-\omega^2 (\omega_{*e} + \omega_{*i}) - \omega\left(k_z^2 c_a^2 - \omega_{*e} \omega_{*i} + k_z^2 c_a^2 k_y^2 c_s^2\right)
%\]
\be
(\omega-\omega_{*e})\left(\omega^2 - \omega_{*i} \omega - k_z^2 c_a^2\right) =  k_z^2 c_a^2  k_y^2
\rho_s^2(\omega-\omega_{*i}).  \label{e8} \ee
This equation is exactly the same as the corresponding two-fluid equation from \citet{vaa}, and it is also obtained
from  the kinetic derivation in \citet{w}.  We stress that such a  perfect agreement between the two (fluid and
kinetic) descriptions is only possible if the two-fluid derivations self-consistently include the gyro-viscosity stress
tensor contributions. Details on these issues can be found  in \citet{w,vaa,vaac}.   From Eq.~(\ref{e8}) it is seen
that the coupling between the drift mode $\omega=\omega_{*e}$ and the Alfv\'{e}n mode is due to the right-hand side,
which here appears due to the finite-ion-mass effect $k_y^2 \rho_s^2$. In the fluid description, the latter term
originates from the ion polarization drift $ \vec v_{ip}=-(\partial/\partial t)[\nabla_\bot \phi_1/(\Omega_i B_0)]$.

\section{Growth rates}
In both  instabilities discussed in the previous section, the perpendicular ion motion is essentially the same: the
typical dominant velocity is due to the $\vec E\times \vec B$-drift and this holds as long as $\lambda_y\gg \rho_i$.
The latter condition may be relaxed, the relative contribution of the polarization drift in that case increases and the
foreseen stochastic heating for the given electrostatic-drift and drift-Alfv\'{e}n instabilities will have a similar
nature. Note that experimental verification performed in \citet{mc2} in fact involved the excitation of the
drift-Alfv\'{e}n waves.    It may be of importance to check the growth rates of these instabilities, calculated for
the same or a similar set of  physical parameters. This may give the answer  about their relative  importance in the
solar corona.

To check our model and the differences between the electrostatic (ES) and electromagnetic (EM) cases,  we take a set of
parameters similar to \citet{v2}: $n_0=10^{15}\;$m$^{-3}$, $B_0=10^{-2}\;$T, and $T_e=T_i=10^6\;$K. We further set
$L_n\equiv [(d n_0/dx)/n_0]^{-1}= s\cdot 10^2\;$m, and take the parallel wave-length $\lambda_z=s\cdot 10^4\;$m. The
plasma $\beta$ for the present case is $0.64 m_e/m_i$; as shown below this  can  be taken as  an appropriate
electrostatic domain.  The parameter $s$ can in principle have any value (e.g.\ in the interval $1-10^3$),  only
bearing  in mind the necessity of staying reasonably well within the previously imposed  conditions used in the
derivations. The simultaneous variation of $L_n$ and $\lambda_z$  by changing the factor $s$ is  introduced for
convenience only: as shown in \citet{v2} for the electrostatic limit, by doing this it turns out that the ratio
$\gamma/\omega_r$ remains exactly the same regardless of the value of $s$, while the actual values of both quantities
$\gamma$ and $\omega_r$ can, in fact, drastically change.

In Fig.~1 we plot the frequency $\omega_r$ for the case $s=1$, and in terms of the perpendicular wavelength
$\lambda_y$.   The four lines in the figure represent the real part of the frequency calculated from Eq.~(\ref{e6}) in
the following manner. The full line is obtained after setting $p=0$, that is equivalent to simply neglecting the EM
effects. In that sense it corresponds to the purely electrostatic analysis given  in \citet{v2}.  The dotted line is
obtained after setting $p=1$, yielding consequently the drift wave frequency when the EM effects are taken into
account. As expected, in view of the given small plasma-$\beta$, the frequency remains almost unchanged. The  frequency
is passing through a maximum, and this  follows from the fact that $\omega_r\sim k_y/(1+ k_y^2 \rho_s^2)$ (note that
$k_y^2 \rho_s^2\simeq 9$ at $\lambda_y=0.2$ and  $k_y^2 \rho_s^2\simeq 0.36$ at $\lambda_y=1$). The  plasma-$\beta$ can
be varied by changing  several parameters. In the present case this is done by the  variation of  the number density.
Hence, we  take it to be $n_0=2\cdot 10^{16}\;$m$^{-3}$ and $n_0=5\cdot 10^{16}\;$m$^{-3}$, and calculate the frequency
from Eq.~(\ref{e6}). This  is represented by the dashed and dash-dotted  lines, respectively.  It is seen that the
drift-wave frequency becomes reduced and this can be attributed  to its coupling with the Alfv\'{e}n wave. Note that
setting  $s=10^3$ (thus  simultaneously changing $L_n$ to $10^2\;$km, and $\lambda_z$ to $10^4\;$km, implying larger
coronal loops) reduces the frequency approximately to $\omega_r/s$, while in the same time the dispersion lines keep
exactly the same shape. Hence, similar to the electrostatic analysis in \citet{v2},  the reduction of frequency due to
variation of $s$ remains more or less the same  even in the presence of EM perturbations  discussed here.

The plot of the growth rate corresponding to the frequencies from Fig.~1, is given in Fig.~2. It shows that the EM
effects can make an important modification of  the drift wave for larger values of the plasma-$\beta$. The reason for
the reduced growth rate  for  short $\lambda_y$ can be understood from Eq.~(\ref{e8}): for larger $k_y$ the coupling
term on the right-hand side becomes more important, a greater amount of  energy is spent on the Alfv\'{e}n mode and,
because of the fixed amount of the free energy stored in the background density gradient,   the growth rate is
therefore reduced. Hence, the physics of the stochastic heating predicted in    \citet{v1,v2,v3,v4} will remain nearly
the same even in the case of the coupling with the Alfv\'{e}n mode, provided a low plasma-$\beta$. On the other hand,
for a larger plasma-$\beta$  the increasing electromagnetic effects will impose longer growth times.

Next, we check the drift mode behavior in terms of the parallel wavelength $\lambda_z$ and this for several different
values of the plasma-$\beta$. For that purpose Eq.~(\ref{e6}) is solved numerically and the results are presented in
Fig.~3 for the plasma number densities $n_0=10^{15}\;$m$^{-3}$, $n_0=6\cdot 10^{15}\;$m$^{-3}$, and
$n_0=10^{16}\;$m$^{-3}$ that yield  $\beta=0.64 m_e/m_i$, $3.8 m_e/m_i$, and $6.4 m_e/m_i$, respectively. Other
parameter values are $\lambda_y=0.5\;$m, $L_n=1\;$km, $T_e=T_i=10^6\;$K. The given shape of the $\gamma/\omega_r$ lines
are equivalent to those from Fig.~3 in  \citet{v2}. Here too, the increased EM effects reduce the growth rate. On the
other hand, similar to  \citet{v2},  for relatively short parallel wavelength components the instability vanishes. This
is due to mobile electrons which now have to move within  shorter distances in the parallel direction in order to
short-circuit the potential buildup caused by the wave.

The plasma-$\beta$ can change also by varying the temperature and/or magnetic field. However, the drift wave frequency
is proportional to the temperature and also strongly depends on the magnetic field [see Eq.~(\ref{k1})], so that the
effect of the perturbed magnetic field alone on the drift-wave in that case is not so transparent. In Figs.~4 and 5 we
give the drift wave frequency and growth rate in terms of the plasma temperature for several values of the plasma
density. Other parameter values are $\lambda_y=0.5\;$m, $\lambda_z=200\;$km, and $L_n=1\;$km. The electromagnetic
effects are again most effectively seen  by taking several values of the plasma density and varying the temperature. At
$T=1.4\cdot 10^6\;$K the frequency is reduced by factor 1.4 for the density increased from $n_0=10^{15}\;$m$^{-3}$ to
$n_0=10^{16}\;$m$^{-3}$. At the same time the corresponding growth rate from Fig.~5 is reduced by a factor 3.2.

The graphs presented in this section are solely for the drift wave part of the spectrum  from Eq.~(\ref{e6}). The
Alfv\'{e}n mode, that is also described by   Eq.~(\ref{e6}), plays no important role in the present study dealing with
the stochastic heating. The coupling of the two modes is in fact given in detail in \citet{vaa} using the two-component
fluid descriptions.

In view of the parameters used in this section, it is seen that the quasi-neutrality condition used in the derivations
is well satisfied. This because the Debye length $\lambda_d$  is typically around $1\;$mm so that $k^2
\lambda_d^2\simeq 0.0004\ll 1$, and in the same time the equivalent condition $\Omega_i^2/\omega_{pi}^2\simeq 0.0005\ll
1$ is also satisfied. For these parameters, the coronal plasma from our examples is rather similar to the tokamak
plasma.

\section{Application to heating}

Considering a specific single coronal loop  and in view of the given geometry that implies a very elongated wave front
in the axial direction, $\lambda_z\gg \lambda_\bot$,  the results presented above  may imply the following. The waves
propagate both axially and poloidally, with drastically different wavelength components in the two directions. In the
cylindric geometry of a magnetic loop, the wave front is thus twisted along the loop.  An extremely large axial
component of the wavelength implies  a wave that simultaneously takes place in areas with gradually varying (with
altitude) plasma parameters and consequently  different   plasma-$\beta$. At higher altitudes with a lower density, the
perturbations may be electrostatic and develop on a shorter time scale. The associated heating may  rapidly develop  at
such places and it can then spread  along the common wave-front   towards lower regions where, due to the increased
plasma-$\beta$, it is additionally accompanied  by the EM effects that develop on  longer characteristic times.

However, this scenario with an increased plasma-$\beta$ due to higher plasma density can be partly counteracted by the
lower temperature at lower altitudes, and as a result the energy release rate and the heating, together with the
variation of  magnetic topology,   may not be so drastically different along the given magnetic loop. One possible
example of this can be seen from Fig.~5: for the given normalized temperature  $T=1$ from the line 1 (point~A in
Fig.~5), the plasma-$\beta$ is $0.64 m_e/m_i$, while for example at $T=0.4$ from the line~3 (point~B), the
plasma-$\beta$ is $2.5 m_e/m_i$. This is of course just a rough estimate because the points A and B belong to two
separate dispersion lines. One particular plasma mode is determined by one particular dispersion line, yet in view of
the parameters changing with the altitude such a transition may be expected.  Hence, the drift wave spreading along the
loop will have an electrostatic nature at the first point, and it will be EM at the second one, implying a difference
in the heating rate and the magnetic variation.

The magnetic field intensity too varies with the altitude, and this may additionally change both the   plasma-$\beta$
and also the wave properties. Hence, assuming the magnetic field is stronger by a  factor 3 (i.e.\ taking $B_0=3 \cdot
10^{-2}\;$T), and for other  parameters as for the line 3 from Figs.~4 and 5, after solving Eq.~(\ref{e6}) again, in
Fig.~6 we present the wave frequency and the growth rate in terms of the temperature.  Observe that, for example, at
the normalized temperature $T=0.4$ (the point~C)  the  plasma-$\beta =0.3 m_e/m_i$, so due to the stronger magnetic
field the mode is now in the electrostatic regime (compare with the point~B from Fig.~5). At the same point we have
the frequency and the growth rate $\omega=13.3 + i 0.22\;$Hz. Hence, for such a stronger magnetic field  the growth
rate becomes  about 50 times lower, as compared to the electrostatic case for the point~A  from Fig.~5.

The points A and C in the previous examples may belong to the same magnetic loop. However,  different loops may have
different plasma-$\beta$ and these points may also represent such a situation. Therefore, the heating in different
loops may be with or without a detectable variation of the magnetic topology. Observations of strong energy release
events  in the past (even in the range of flares) have shown that both scenarios are indeed possible; examples without
magnetic variations can be seen   in \citet{jan, mc, pud}. The  qualitative analysis described above  will additionally
be supported below  by some  more numbers.

According to \citet{b1, mc2, b2} the stochastic heating by the drift wave is in action provided a strong enough wave
potential amplitude, more precisely if
\be
a\equiv \frac{m_i k_y^2 \phi_1}{e B_0^2}=  k_y^2 \rho_i^2 \cdot \frac{e \phi_1 }{\kappa T_{i0}} \geq 1, \label{h2} \ee
and the maximum achieved ion velocity  due to this heating is given by
 \be
  v_{max}\simeq [k_y^2 \rho_i^2 e
\phi_1/(\kappa T_i) + 1.9]\Omega_i/k_y. \label{vm} \ee
The effective stochastic temperature is then  $T_{max}= m_i v_{max}^2/(3 \kappa)$. From Eq.~(\ref{h2}) it follows that
the condition for the stochastic heating will sooner be satisfied  in regions of a weaker magnetic field, regardless of
the starting temperature. The physics of the heating is described in \citet{b1, mc2, b2} and its electrostatic
application to the solar corona  in \citet{v1,v2,v3,v4}, so the details of this will not be repeated here. We shall
only stress that the heating is essentially due to the ion polarization drift.  We shall apply these expressions using
the ES and  EM  growth rates given above in order to quantitatively verify the differences due to eventual magnetic
nature of the perturbations.

For the parameters corresponding to the point~A from Fig.~5, the expression (\ref{h2}) for $a=1$  yields the required
potential $\phi_1=61\;$V. The maximum achieved stochastic velocity from Eq.~(\ref{vm}) is $221\;$km/s, and the achieved
stochastic temperature is $T_{max}=1.97\cdot 10^6\;$K.  Assuming some small accidental initial perturbations with the
amplitude  $e\hat{\phi}/(\kappa T_i)=0.01$, i.e., $ \hat{\phi}=0.86\;$V we can calculate the time $t_g$ required to
achieve the required value for the heating $\phi_1=\hat{\phi} \exp( \gamma t_g)$. This yields $t_g=\ln
(\phi_1/\hat{\phi})/\gamma=0.4\;$s. Note also that here $e \phi_1/(\kappa T_i)\simeq 0.7$.   The total released energy
density is $E_{max}=n_0 m_i v_{max}^2/2=0.04\;$J/m$^3$, and the energy release rate
$\Gamma_{max}=E_{max}/t_g=0.1\;$J/(m$^3$s). Hence, $\Gamma_{max}$ is about 1700 times the required value for the
coronal active regions [that amounts to $\simeq 6 \cdot 10^{-5}\;$J/(m$^3$s) \citep{nar}].

Taking as another example  the point~B in Fig.~5 (i.e., the same magnetic field   $10^{-2}\;$T but different density
and starting temperature) yields  $E_{max}=0.4\;$J/m$^3$, $t_g=1.3\;$s, and $\Gamma_{max}=0.3\;$J/(m$^3$s),
$T_{max}=1.97\cdot 10^6\;$K; $\beta=2.5 m_e/m_i$. So the present case is weakly electromagnetic and it is accompanied
with   the increase in the energy density  and the energy release rate (because the density is higher), as compared to
the electrostatic case from the point~A, although it implies a longer growth time. The achieved stochastic temperature
$T_{max}$ and velocity $v_{max}$  are the same as in the point~A because $a$ is kept fixed, and $\phi_1=61\;$V $,
\hat{\phi}=0.34\;$V as in the previous case.

On the other hand, taking  as example the point~C from Fig.~6,  and the threshold (\ref{h2}),  yields:  $
\hat{\phi}=0.34\;$V,  $\phi_1=546\;$V,  $v_{max}= 663\;$km/s, $T_{max}=1.8\cdot 10^7\;$K, $E_{max}=3.67\;$J/m$^3$,
$t_g=33\;$s, and $\Gamma_{max}=0.11\;$J/(m$^3$s). The stronger necessary potential here is due to the increased value
of the magnetic field, see Eq.~(\ref{h2}). Hence, the energy release rate $\Gamma_{max}$ is almost the same as for the
point~A, yet the characteristic time $t_g$ for the point~A is more than 80 times shorter. In the same time, the maximum
released energy density in the area with such a stronger  magnetic field $B_0$  is for about a  factor 90  larger in
comparison with the point~A, with  the achieved stochastic temperature that goes to 18 million K. The reason for the
larger energy density is clearly the larger maximum stochastic velocity in the area where both the magnetic field and
density are larger. The fact that plasma-$\beta$ for this stronger magnetic field  is only $0.3 m_e/m_i$ tells us that
the increased stochastic energy density can be related to the EM effects and the coupling with the Alfv\'{e}n wave.
However, a much more pronounced effect on the heating  should be attributed  to the increased  magnetization of the
plasma species. Thus, the areas with  stronger background  magnetic fields are subject to stronger stochastic heating.
The magnetic field used here is in agreement with observations  of active regions showing the field strength of a few
times $0.01\;$T, that in fact may easily go above $0.1\;$T \citep{lee,sol}, implying a possibly still stronger heating
within the scenario presented above.

Note also that the two potentials for the points A and C, $61\;$V and $546\;$V, respectively, are obtained assuming
$a=1$ in Eq.~(\ref{h2}). These  two potentials  yield  the electric field $k_y \phi_1$  in the perpendicular direction
$0.77\;$kV/m and $6.9\;$kV/m, respectively.   Now, to have the threshold $a=1$, the required potential $\phi_1\sim
\lambda_y B_0^2$, so a slight increase in these two parameters will yield even  stronger electric fields. Taking as
example $\lambda_y=2\;$m, $B_0=4\cdot 10^{-2}$ (instead of $\lambda_y=0.5\;$m, $B_0=3\cdot 10^{-2}$ as in the point~C)
yields the perpendicular electric field at which the stochastic heating takes place $E_y\simeq 27\;$kV/m.  The three
obtained  values for the electric field, together  with the corresponding magnetic field values, yield the $\vec
E\times\vec B$-drift ($=E/B_0$) of the plasma as a whole in the perpendicular direction $77$, $230$, and $675\;$km/s,
respectively. In view of the meter-sized perpendicular wavelengths these plasma flows (drifts) could eventually be
observed only by spectral analysis. Hence, we conclude that i) exceptionally  strong perpendicular electric fields are
expected during the proposed stochastic heating, and this particularly within stronger magnetic structures, and, ii)
the perpendicular stochastic heating, as single particle interaction with the wave, is accompanied with collective
plasma drifts.

\section{Summary and conclusions}

The results presented in this  work could be summarized as follows. The  kinetic theory of the drift wave shows that
the mode is almost always  unstable due to purely kinetic effects, and it couples naturally  to the Alfv\'{e}n wave.
The higher the plasma-$\beta$ is,  the better the coupling. The electrostatic drift wave in the solar corona is
expected to be more unstable as compared to the regime in which the two modes are coupled. Essential for plasma heating
is the electrostatic part of such an electromagnetic drift-Alfv\'{e}n wave. The heating is stochastic by nature and, as
shown in our previous works \citet{v1, v2, v3, v4}, it  possesses  such properties that it is able to satisfy numerous
heating requirements in the solar corona. From the analysis it also follows that the regions with stronger magnetic
fields will be subject to much stronger heating. Note that such a relation between the temperature and the magnetic
field has been established long ago \citep{vs}; in the present work we  give a new,  alternative explanation for this
phenomenon.

The electric field associated with the drift wave implies the possibility of the acceleration of plasma particles
(primarily electrons) in the direction  parallel to the magnetic field vector, and        the development of drifts (in
the perpendicular direction)  of the plasma as a whole due to the $\vec E\times \vec B$-drift that is the same for both
electrons and protons. This issue is discussed in \citet{v2, v3}. The mean free path of the plasma species $j$ is
proportional to $\vtj^4$ and therefore the parallel acceleration by the electric field is always more effective on
particles that are already faster, i.e., those from the tail in the starting (Maxwellian) distribution, because those
have more time/space to interact with the field.  This will consequently  result in a very different distribution
function with a  much longer high-velocity tail and resembling the $\kappa$-distribution observed in the outer solar
atmosphere. The proposed electron acceleration within the present model appears as a natural development of the drift
wave instability for which the source is clearly identified, thus removing the standard problem of various acceleration
schemes that typically suffer from a common problem, the lack of a proper source.

Some phenomena that follow from the presented stochastic heating are not discussed here, but they are  given in detail
in \citet{v2}. These include the fact that the proposed model explains the better heating of heavier particles (i.e.,
heavier ions are better heated than lighter ones, while the ions in general are better heated than electrons). This
follows after analyzing the mass dependence of the stochastic temperature $T_{max}$ introduced earlier, with the help
of Eq.~(\ref{vm}). Also the better heating in the direction perpendicular to the magnetic field vector, and the
associated temperature anisotropy $T_\bot> T_{\|}$,  is  self-evident and explained in \citet{v2} as a consequence of
the polarization drift that acts primarily in the perpendicular direction and becomes important at perpendicular
wavelengths close to the ion gyro radius.

The value of $e \phi_1/(\kappa T_i)$ in general determines the importance of nonlinearities. It turns out that in the
examples discussed in the text this quantity is not small so that the presented scenario, which follows from the linear
theory,  may change considerably. In addition, the effects of  nonlinearity in the drift-wave theory are determined
also by making the ratio of the nonlinear term (i.e., the convective derivative in the momentum equation), and the
leading order linear term \citep{has}. The result can be written as  $ (k_y L_n) ( k_y^2 \rho_s^2) [e\phi_1/(\kappa
T)]= k_yL_n a$. Because $L_n\gg \lambda_y$, the proposed stochastic heating will be accompanied by various   nonlinear
phenomena. The most important  nonlinear effects expected here  include  nonlinear 3-wave interaction that implies the
well known double cascade (transfer of energy of a large amplitude drift-wave towards both longer and shorter
wavelengths), and the anomalous transport caused by drift wave turbulence. These effects, however important, require
numerical simulations and will be studied elsewhere.

\acknowledgments

These results were obtained in the framework of the projects GOA/2009-009 (K.U.Leuven), G.0304.07 (FWO-Vlaanderen) and
C~90347 (ESA Prodex 9). Financial support by the European Commission through the SOLAIRE Network (MTRN-CT-2006-035484)
 is gratefully acknowledged.

\pagebreak

\clearpage

\begin{figure}
\includegraphics[height=10cm, bb=15 14 275 215,clip=,width=.9\columnwidth]{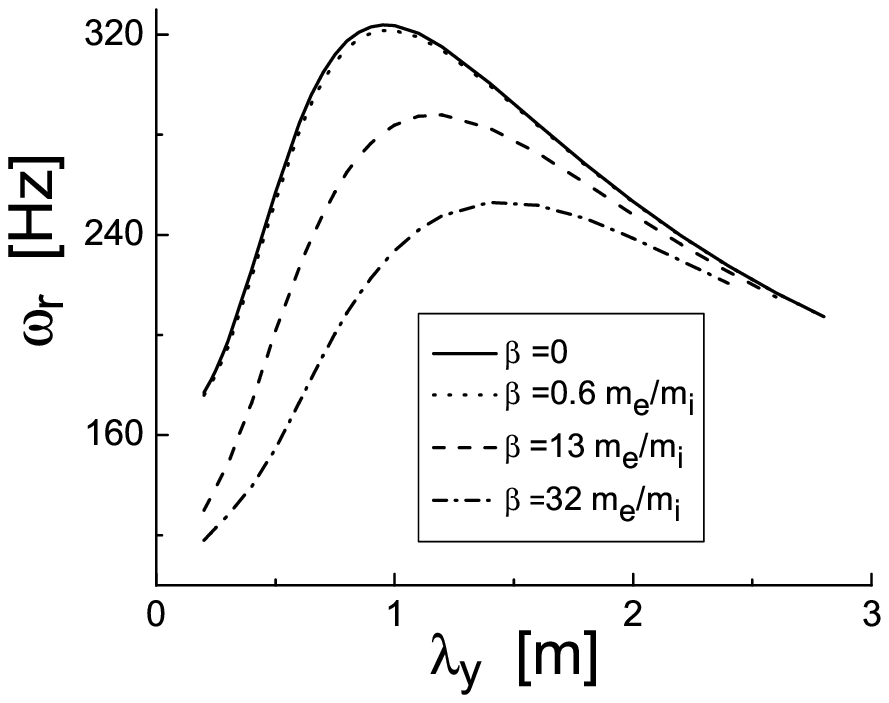}
%\vspace*{-5mm}
 \caption{The drift wave frequency for several  different values of plasma $\beta$ in terms of the perpendicular wave-length. }\label{fig1}
% \vspace{0.3cm}
\end{figure}

\clearpage

\begin{figure}
\includegraphics[height=10cm, bb=10 10 300 230,clip=,width=.9\columnwidth]{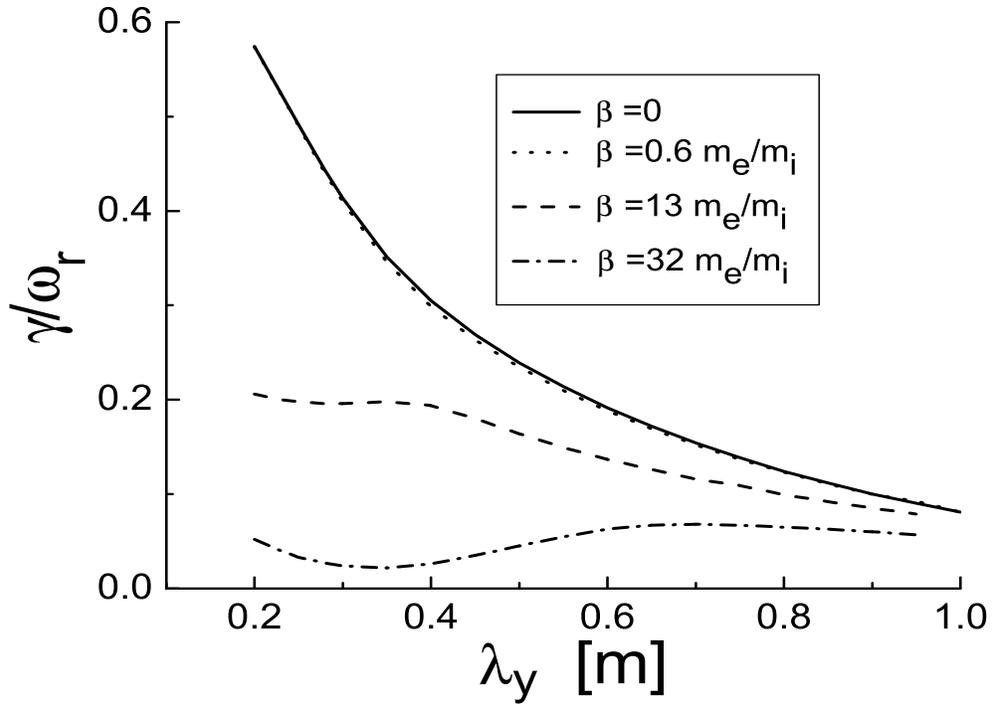}
%\vspace*{-5mm}
 \caption{The normalized growth rates corresponding to the frequencies from Fig.~1.}\label{fig2}
% \vspace{0.3cm}
\end{figure}

\clearpage

\begin{figure}
\includegraphics[height=10cm, bb=10 10 280 225,clip=,width=.9\columnwidth]{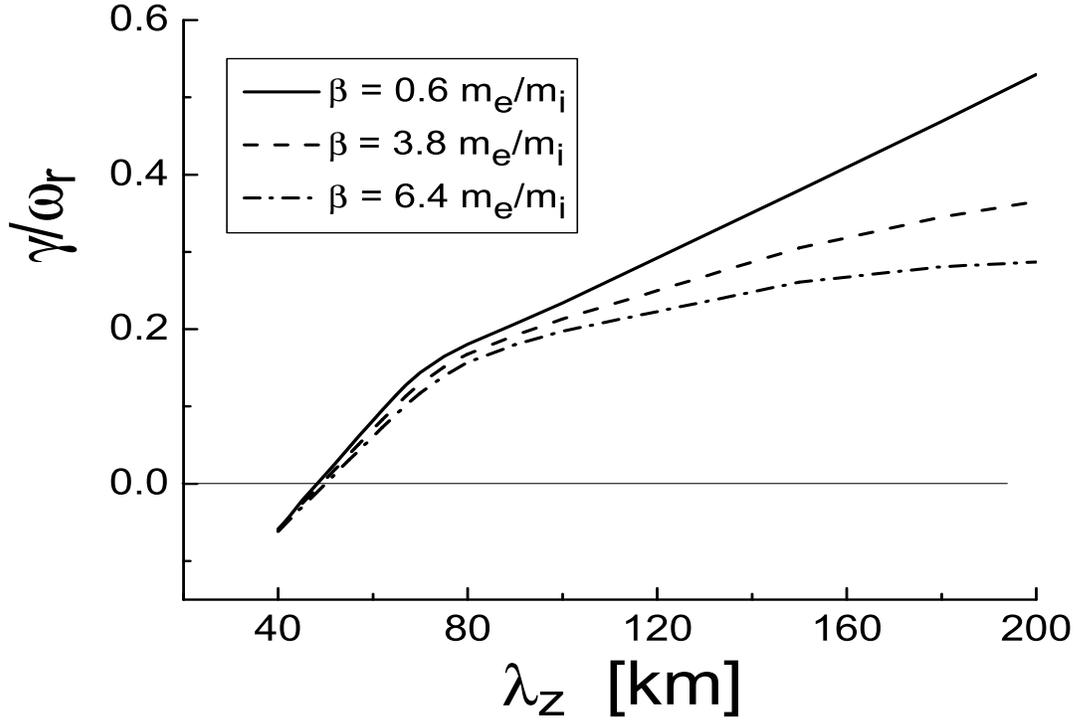}
%\vspace*{-5mm}
 \caption{The normalized growth rate of the drift-wave obtained from Eq. (\ref{e6}) in terms of the parallel wavelength and for three plasma number densities $n_0=10^{15}$ m$^{-3}$ (full line),
 $n_0=6 \cdot 10^{15}$ m$^{-3}$ (dashed line), $n_0=10^{16}$ m$^{-3}$ (dashed-dotted  line).  }\label{fig3}
% \vspace{0.3cm}
\end{figure}

\clearpage

\begin{figure}
\includegraphics[height=10cm, bb=10 10 280 225,clip=,width=.9\columnwidth]{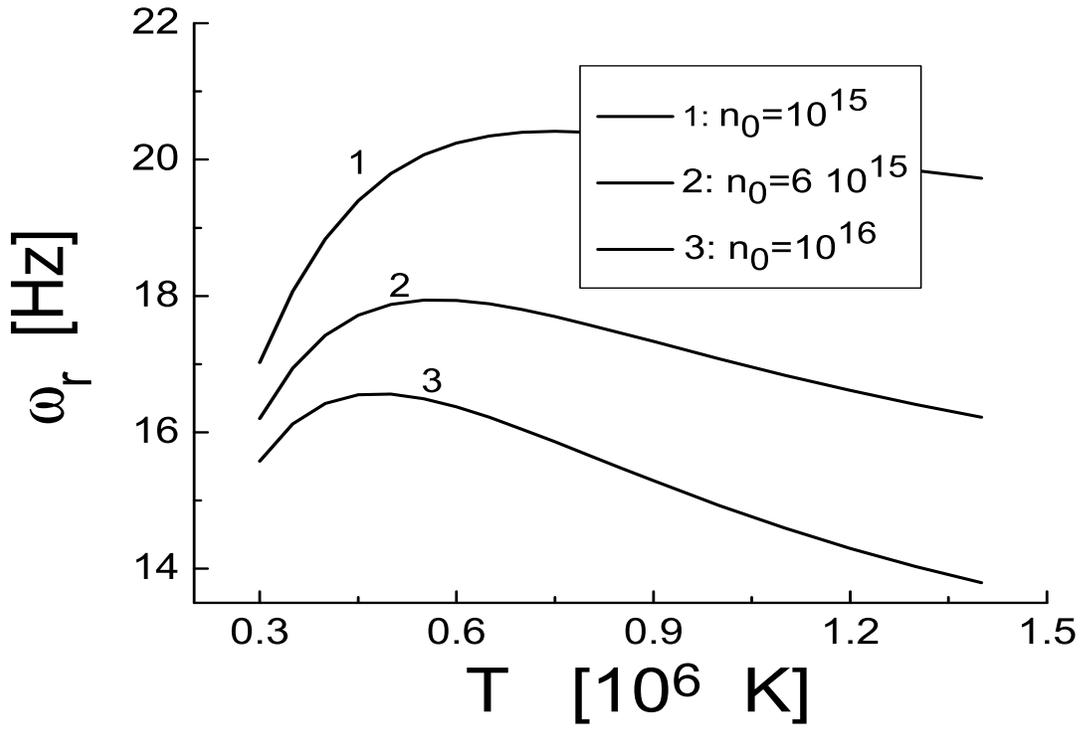}
%\vspace*{-5mm}
 \caption{The drift wave frequency  obtained from Eq. (\ref{e6}) in terms of the plasma temperature and for several values of the plasma number density (per cubic meter).  }\label{fig4}
% \vspace{0.3cm}
\end{figure}

\clearpage

\begin{figure}
\includegraphics[height=10cm, bb=10 10 280 225,clip=,width=.9\columnwidth]{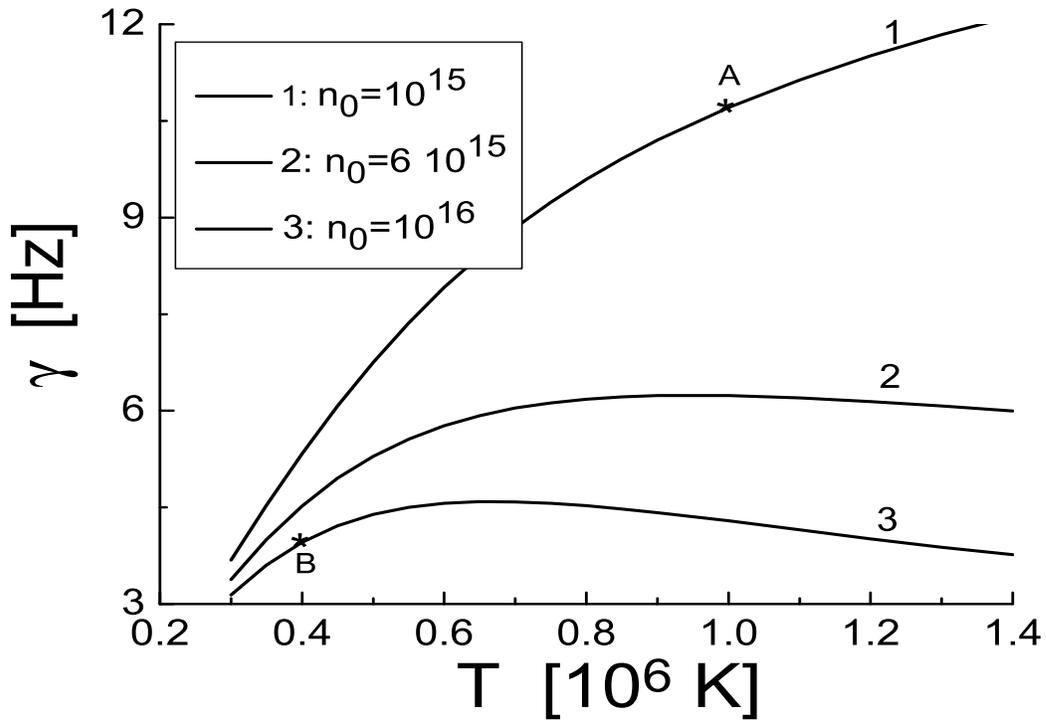}
%\vspace*{-5mm}
 \caption{The drift wave growth rate corresponding to the frequency from Fig.~4.    }\label{fig5}
% \vspace{0.3cm}
\end{figure}

\clearpage

\begin{figure}
\includegraphics[height=10cm, bb=10 10 280 225,clip=,width=.9\columnwidth]{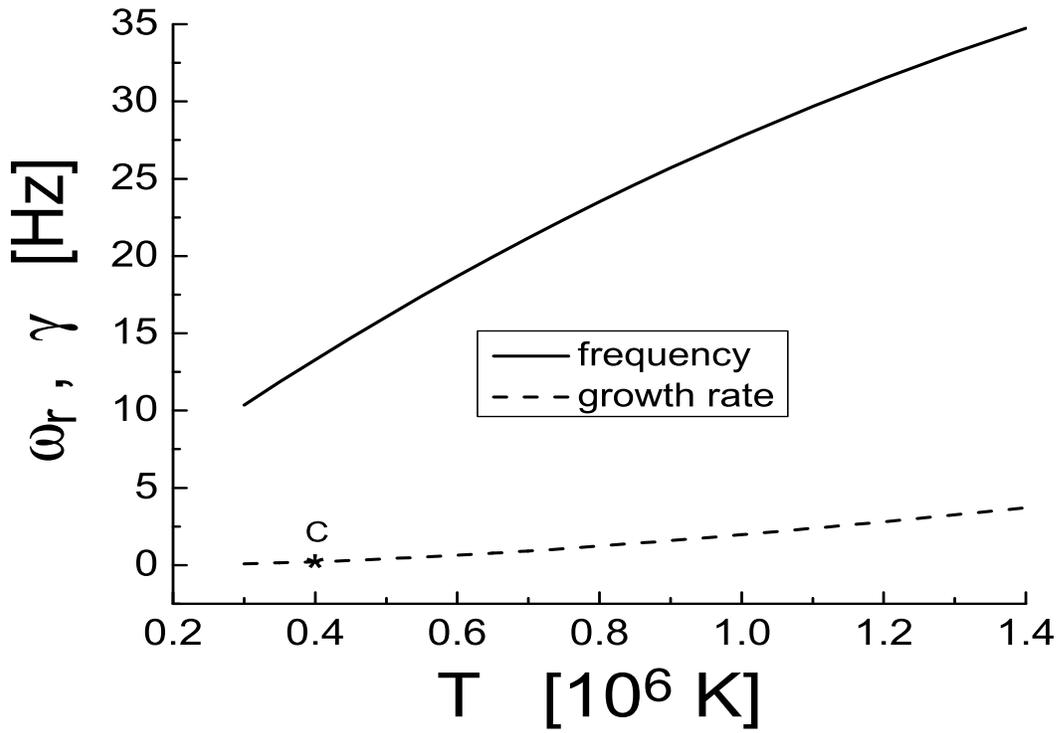}
%\vspace*{-5mm}
 \caption{The drift wave frequency and the growth rate for $B_0=3\cdot 10^{-2}$ T; other parameters are the same as for line 3  from Figs.~4, 5.    }\label{fig6}
% \vspace{0.3cm}
\end{figure}

\clearpage

\end{document}